\documentclass{article}

\PassOptionsToPackage{numbers, compress}{natbib}

\usepackage[final]{neurips_2022}

\usepackage[utf8]{inputenc} 
\usepackage[T1]{fontenc}    
\usepackage{hyperref}       
\usepackage{url}            
\usepackage{booktabs}       
\usepackage{amsfonts}       
\usepackage{nicefrac}       
\usepackage{microtype}      
\usepackage{xcolor}         
\usepackage{graphicx}       
\usepackage{multirow}       

\graphicspath{ {images/} }

\title{Instrument Separation of Symbolic Music \protect\\ by Explicitly Guided Diffusion Model}

\author{
  Sangjun Han \\
  LG AI Research \\
  \texttt{sj.han@lgresearch.ai} \\
  \And
  Hyeongrae Ihm \\
  LG AI Research \\
  \texttt{hrim@lgresearch.ai} \\
  \AND
  DaeHan Ahn \\
  University of Ulsan \\
  \texttt{daehan.ahn@ulsan.ac.kr} \\
  \And
  Woohyung Lim \\
  LG AI Research \\
  \texttt{w.lim@lgresearch.ai} \\
}

\begin{document}

\maketitle

\begin{abstract}
  Similar to colorization in computer vision, instrument separation is to assign instrument labels (e.g. piano, guitar...) to notes from unlabeled mixtures which contain only performance information.
  To address the problem, we adopt diffusion models and explicitly guide them to preserve consistency between mixtures and music.
  The quantitative results show that our proposed model can generate high-fidelity samples for multitrack symbolic music with creativity.
\end{abstract}

\section{Introduction}

With the recent progress of music technology, a musician can play multiple instruments simultaneously by dividing pitches of a electronic keyboard into several zones to deploy each instrument.
This feature is known as \textit{zoning}, however, it has limitations for the solo ensemble because of limited and overlapped pitch ranges for each instrument.
To support musicians during their solo performance, Dong et al. \citep{dong2021towards} have introduced \textit{instrument separation} or \textit{Mixture2Music} to assign the instruments dynamically to notes in the ensemble.
As colorization in computer vision, it is placed on an ill-posed problem since multiple instruments can be labeled on single note.

The solution for the above problem requires probabilistic models or additional constraints appropriate for music.
First, the development of diffusion models has shown promising results for generation tasks, including music generation \citep{mittal2021symbolic}, only assuming the type of noise processes with the Markov property. This design can enable to deal with \textit{Mixture2Music} effectively without any differentiable tricks for binary pianoroll since it has few architectural assumptions \citep{dong2018musegan, dong2018convolutional}.
Second, one of the constraints from \textit{Mixture2Music} is on how to maintain strong consistency between mixtures and generated music.
It means that the task should satisfy the condition that all notes in any generated music belong to one of the components in a mixture.
Although recent diffusion models suggest the ways of guiding directions conditioned on labels \citep{dhariwal2021diffusion, ho2022classifier}, it just only guarantees weak constraints.

In this paper, we utilize diffusion models to address the instrument separation and impose a consistency constraint on them.
For feasibility experiments, mixtures that are solo performance generated virtually from the multitrack music. Our results are evaluated for creativity as well as consistency.

\section{Method}

In this section, we introduce the preparation process of our dataset, the special point of the instrument separation model, and its evaluation protocol.

\begin{figure}
  \centering
  \includegraphics[width=\textwidth]{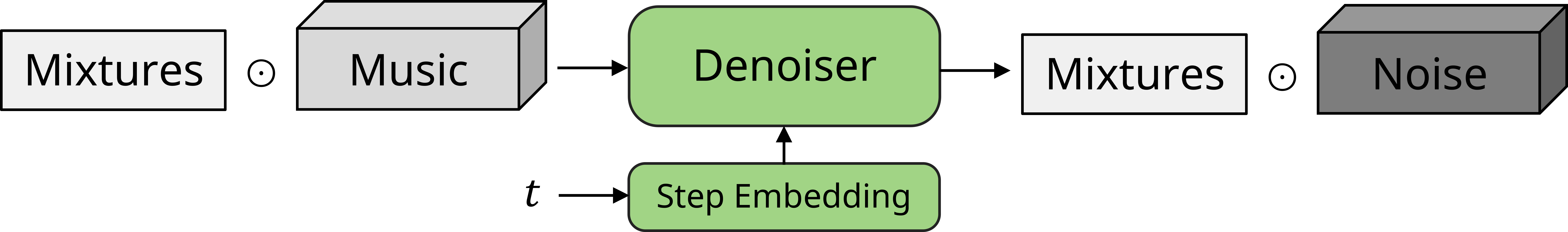}
  \caption{The denoiser for instrument separation.}
\end{figure}

\subsection{Data preparation}

For symbolic music generation, Lakh MIDI Dataset \citep{raffel2016learning} is the most prevalent because it contains 176,581 MIDI with various genres and tracks.
We take only 4/4 time signature MIDIs which consist of more than one instrument that we defined.
The tracks in each MIDI can be composed of the combinations of piano, guitar, bass, string, and drum.
To convert the MIDIs to pianoroll representation, we adopt 16th note quantization with binary representation.
By a sliding window with a stride of 1 bar, we obtained 3,106,506 music phrases each of which can be denoted as \(y^{T\times P\times C}\in \left\{0, 1\right\}\) where \(T\) indicates the number of time steps in a bar (\(T=64\)), \(P\) for the number of pitches (\(P=72\)), and \(C\) for the number of instruments (\(C=5\)).

From the acquired music phrases, mixtures \(x^{T\times P}\in \left\{0, 1\right\}\) can be generated by summing and clipping \(y\) along \(C\) dimension. Then, our instrument separation task is defined to recover \(y\) from \(x\).

\subsection{Separator based on diffusion model}

Our separator is built on DDPM settings \citep{ho2020denoising} with forward and reverse processes of Gaussian transformation. Unlike \citep{mittal2021symbolic}, we diffuse the pianoroll \(y\) directly without any pre-trained models.
The simplified training objective from DDPM, which predicts noise at time step \(t\), is also adopted with linear noise scheduling for 1,000 time steps. 
To fulfill model generalization and large capacity, TransUNet \citep{chen2021transunet} which adopts optimal fusion of convolutions and attentions is used as a denoiser and modified to admit time step embeddings through adaptive group normalization layers (Figure 1).

When training the model, a mixture is conditioned by being multiplied to both the inputs and targets of the denoiser. This is an effective strategy because zero indicators in the mixture prevent generated music from activating notes for all instruments. It enables to preserve strong consistency between mixtures and music. At the inference phase, the reverse process starting from isotropic Gaussian noise is conducted, maintaining mixture multiplications for all time steps. To obtain sigmoid probability of each note at the last time step, we train another simple decoder that denoises noisy samples from \(t=1\) to \(t=0\).

\section{Evaluation}

We suggest two metrics; 1) consistency that measures how similar mixtures (\(x^{'}\)) obtained from generated samples are to their original mixtures (\(x\)), 2) diversity that measures the average of distance between generated samples (\(y^{'}\)) and their original music (\(y\)). All distances are computed by the average of element-wise Hamming distance.

We compare three models based on TransUNet with similar scale; 1) VAE with mixture mask at the end of layers, 2) DDPM, and 3) DDIM \citep{song2020denoising}. Compared to VAE (Table 1), the diffusion models tend to generate more consistent and diverse samples for instrument separation. Our results also indicate that the number of diffusion time steps is sufficient since the DDPM is comparable to the DDIM.

\begin{table}[b]
  \caption{The evaluation table. Best values are marked in bold font.}
  \centering
  \begin{tabular}{lll}
    \toprule
    Model     & Consistency (\(\downarrow\))    & Diversity (\(\uparrow\)) \\
    \midrule
    VAE       & 4.500e-3                        & 8.625e-3 \\
    DDPM      & \textbf{8.974e-5}               & 1.107-2 \\
    DDIM      & 2.469e-4                        & \textbf{1.138e-2} \\
    \bottomrule
  \end{tabular}
\end{table}

\section{Ethical Implications}

Generated contents from our proposed model are not new but rearranged versions of the existing music. Original creator of Lakh MIDI Dataset can claim his copyright.

\bibliographystyle{unsrtnat}
\bibliography{ref}

\newpage

\appendix

\section{Visualization for DDIM Results}

We introduce three examples of instrument separation; the case of 1) recovering like its original music, 2) missing some tracks, and 3) over-generating new tracks. For each figure, the upper one indicates a generated sample from a mixture and the bottom one indicates the original music of the mixture. From left to right, each distinct color area denotes the instrument set (piano, guitar, bass, string, and drum). You can listen more samples at https://github.com/sjhan91/Mixture2Music\_Official.

\begin{figure}[htb!]
  \centering
  \includegraphics[width=0.8\textwidth]{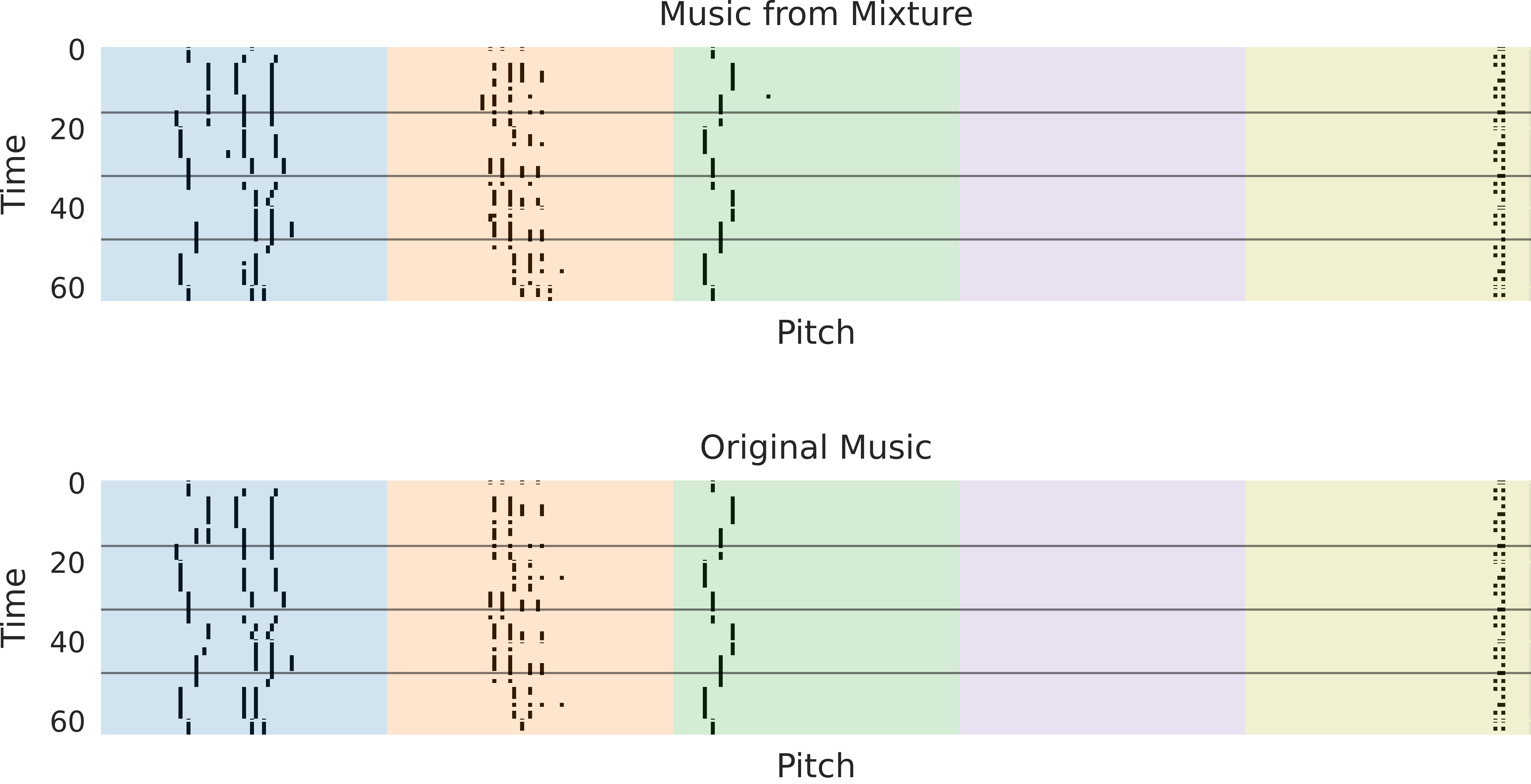}
\end{figure}
\hrule
\begin{figure}[htb!]
  \centering
  \includegraphics[width=0.8\textwidth]{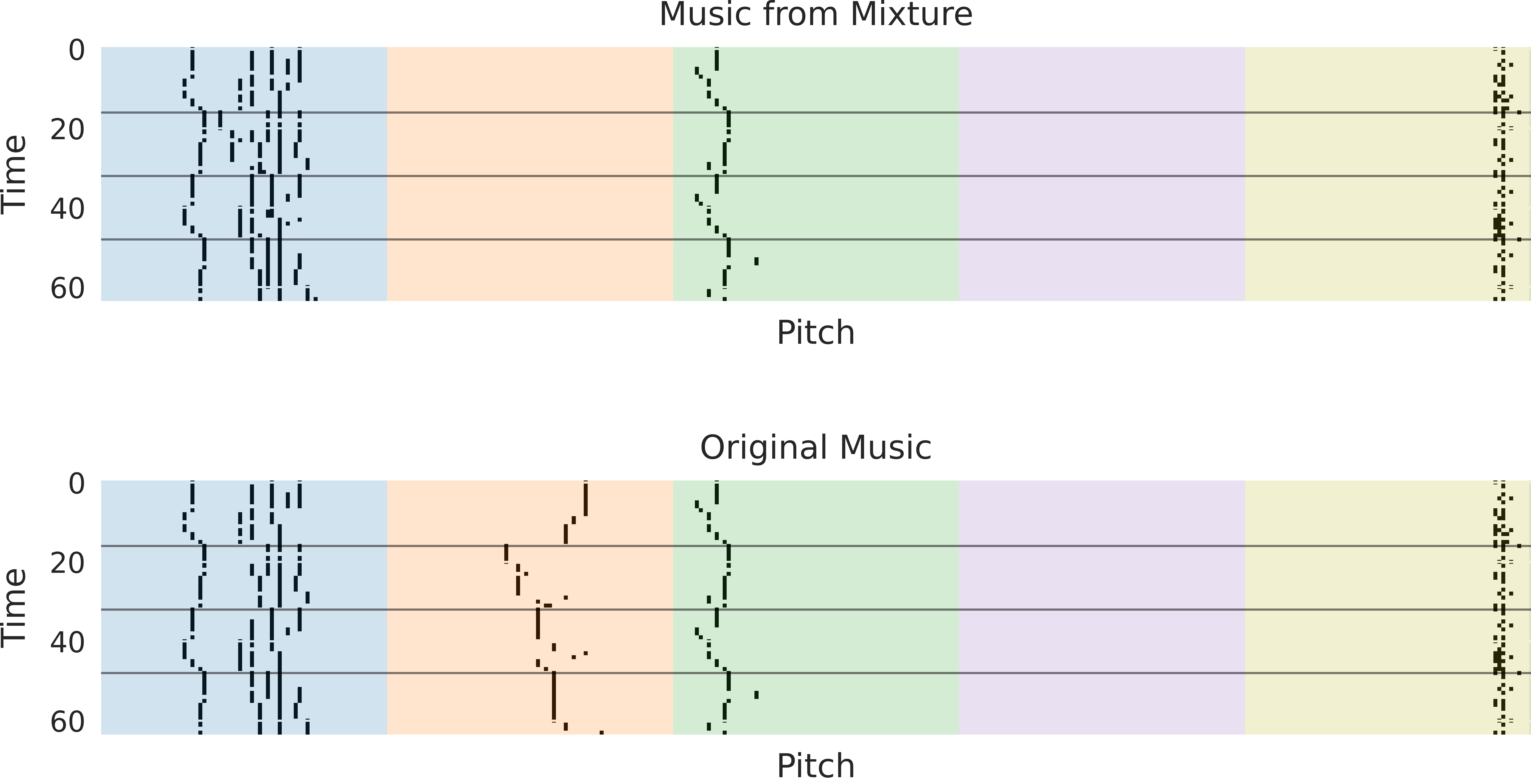}
\end{figure}
\hrule
\begin{figure}[htb!]
  \centering
  \includegraphics[width=0.8\textwidth]{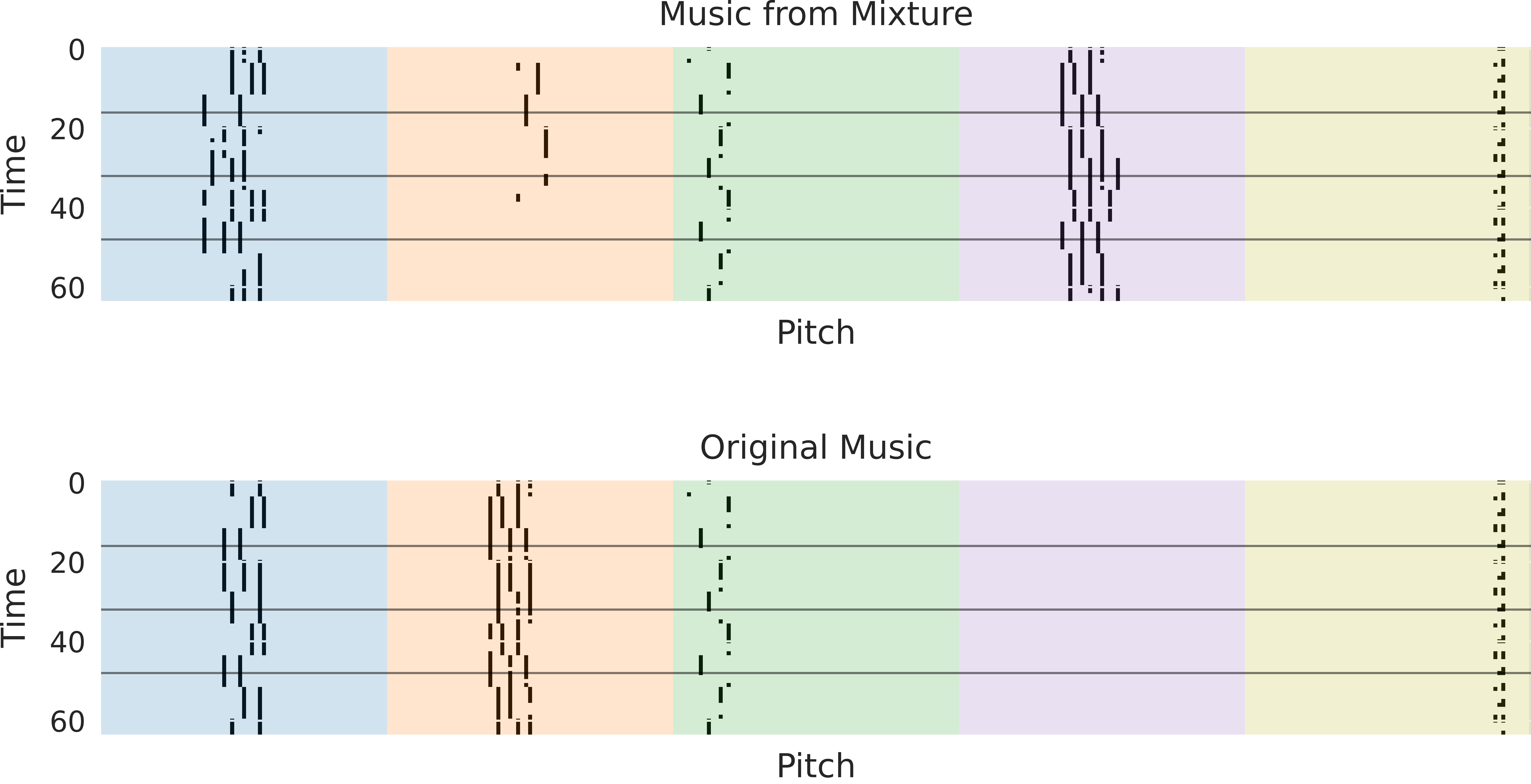}
\end{figure}

\newpage

\section{Implementation Details}

You can refer to the structure of our denoiser derived from TransUNet in Table 2. The down-block contains max-pooling and residual convolution layers, and the up-block does interpolation, concatenation of encoded features, and residual convolutions. In the bottleneck of the denoiser, there are 2 Transformer layers each of which has 256 hidden units with 4 attention heads. Their positional encodings are learned in a relative manner.

\begin{table}[h]
    \caption{Our TransUNet structure. The input and output shape are (\(T=64\), \(P=72\), \(C=5\)).}
    \centering
    \begin{tabular}{ccc}
        \toprule
        Model                    & Operations   & Output Shape  \\
        \midrule
                                 & Input Layer  & (64, 72, 64)  \\
        \multirow{3}{*}{Encoder} & Down         & (32, 36, 128) \\
                                 & Down         & (16, 18, 256) \\
                                 & Down         & (8, 9, 256)   \\
                                 & Transformer  & (8, 9, 256)   \\
        \midrule
        \multirow{4}{*}{Decoder} & Up           & (16, 18, 128) \\
                                 & Up           & (32, 36, 64)  \\
                                 & Up           & (64, 72, 64)  \\
                                 & Output Layer & (64, 72, 5)   \\
        \bottomrule
    \end{tabular}
\end{table}

Time embeddings are expanded through sinusoidal positional embeddings (16 dims) and conditioned on adaptive group normalization in the convolution layers. Our model is trained on 50 epochs with AdamW optimizer, reducing learning rates (starting from 1e-3) by a factor of 0.9 when the validation error does not increase.

For the VAE baseline implementation, bottleneck features of the TransUNet (after Transformer layers) are reparameterized to be sampled from Gaussian distribution. It means that the VAE has larger parameters than the diffusion models because of the layers responsible for mean and variance vectors.

\end{document}